\documentclass[a4paper]{jpconf}
\usepackage{graphicx}
\usepackage{graphics}
\usepackage{epsfig}
\usepackage{bbm}
\usepackage{amssymb}

\newcommand{\be}{\begin{equation}}
\newcommand{\ee}{\end{equation}}
\newcommand{\eea}{\end{eqnarray}}
\newcommand{\bea}{\begin{eqnarray}}

\newcommand{\mean}[1]{\ensuremath{\langle{#1}\rangle}}

\newcommand{\eins}{\ensuremath{\mathbbm 1}}
\newcommand{\II}{\ensuremath{\mathbbm I}}
\newcommand{\qed}{\ensuremath{\hfill \Box}}

\newcommand{\WW}{\ensuremath{\mathcal{W}}}
\newcommand{\FF}{\ensuremath{\mathcal{F}}}
\newcommand{\GG}{\ensuremath{\mathcal{G}}}
\renewcommand{\NL}{\ensuremath{\mathcal{X}}}

\newcommand{\HH}{\ensuremath{\mathcal{H}}}
\newcommand{\BB}{\ensuremath{\mathcal{B}}}

\newcommand{\ketbra}[1]{\ensuremath{| #1 \rangle \langle #1 |}}

\newcommand{\ket}[1]{\ensuremath{|#1\rangle}}

\newcommand{\bra}[1]{\ensuremath{\langle#1|}}

\newcommand{\braket}[2]{\ensuremath{\langle #1|#2\rangle}}

\newcommand{\kommentar}[1]{}
\newcommand{\vr}{\ensuremath{\varrho}}
\begin{document}
\title{Nonlinear entanglement witnesses, covariance 
matrices and the geometry of separable states}

\author{Otfried G\"uhne}

\address{Institut f\"ur Quantenoptik und Quanteninformation,
\"Osterreichische Akademie der Wissenschaften, A-6020 Innsbruck,
Austria}
\ead{otfried.guehne@uibk.ac.at}

\author{Norbert L\"utkenhaus}
\address{Institute for Quantum Computing and Department of Physics 
and Astronomy, University of Waterloo,
200 University Avenue West, Waterloo, Ontario N2L 3G1, Canada}

\ead{nlutkenhaus@iqc.ca}

\begin{abstract}
Entanglement witnesses provide a standard tool for the analysis
of entanglement in experiments. We investigate possible nonlinear 
entanglement witnesses from several perspectives. First, we 
demonstrate that they can be used to show that the set of 
separable states has no facets. Second, we give a new derivation 
of nonlinear witnesses based on covariance matrices. Finally, we 
investigate extensions to the multipartite case.
\end{abstract}

\section{Introduction}
Entanglement plays an outstanding role in many 
protocols of quantum information theory. 
Consequently it is under intensive research from
several perspectives: from the theoretical side
many efforts are undertaken to recognize it via 
separability criteria \cite{ppt1, ppt2, ccn1, ccn2, cmc} or to quantify it 
via entanglement measures \cite{plenio}. {From} the experimental
side, a huge amount of work is devoted to the experimental
generation of entanglement using photons \cite{mohamed, zeil, lu}, ions in a 
trap \cite{ions2, ions1}, or solid state systems 
\cite{solids}.

To confirm the success of such an experiment, one has to 
verify that entanglement was indeed produced. Here it is 
important to perform the analysis without making use of 
hidden assumptions concerning the state \cite{enk}.
Entanglement witnesses are a versatile tool for 
this entanglement verification 
\cite{ppt2, mohamed, lu, terhal, optimization, altpra}. 
They are observables which have, by construction, a positive 
expectation value on all separable states, hence a negative 
expectation value signals the presence of entanglement. 
Besides the mere detection, they also allow for a 
quantitative analysis by giving bounds on entanglement 
measures \cite{werner, jens}. Consequently, they are now 
widely used in experiments. 

In Ref.~\cite{wir} it has been shown that one can improve all
entanglement witnesses for bipartite systems by nonlinear 
correction terms. In 
this paper we extend the analysis of these nonlinear 
entanglement witnesses in several directions. First, in Section 2, 
we recall some facts concerning entanglement and entanglement 
witnesses. We explain the underlying definitions, their 
geometrical interpretation and the main idea for nonlinear 
witnesses. In Section 3 we show how nonlinear witnesses 
can be constructed starting from any witness for bipartite systems. 
We follow the proof  from Ref.~\cite{wir} here and use it to 
derive that the set of separable states has no facets. 
In Section 4 we give an alternative proof for the fact that any 
entanglement witness can be improved. This proof uses
covariance matrices of special observables for the derivation, 
highlighting the close connection between the theory of nonlinear
witnesses and separability criteria in terms of covariance 
matrices \cite{cmc, vogel, rigas, piani}. Finally, in Section 5 we discuss to which extent
the presented methods may be used to derive nonlinear entanglement 
witnesses for the multipartite case.

\section{Separability and the idea of nonlinear witnesses}

Let us first recall the definition of entanglement and 
separability \cite{rfw}. By definition, a quantum state 
$\vr$ shared between Alice and Bob is separable if it can 
be written as a mixture of product states, that is
\be
\vr= \sum_{i} p_i \ketbra{a_i} \otimes \ketbra{b_i},
\label{sepdef}
\ee
where $p_i \geq 0$ and $\sum_i p_i =1,$ that is, the $p_i$ 
form a probability distribution. If this is not the case, 
then $\vr$ is called entangled.

Physically, this definition means that a separable state can 
be produced using local operations and classical communication: 
by public communication, Alice and Bob can agree on producing 
the states $\ketbra{a_i} \otimes \ketbra{b_i}$ locally and agree 
on the probabilities $p_i$ for these states. 

Geometrically, this definition implies that the set of separable 
states is a convex set. First, the set of all states, that is, 
all positive matrices with $Tr(\vr)=1$ is a convex set. Then 
the set of separable states is a convex subset. It has the 
product states as its extremal points, since any separable
state can be written as a convex combination of product states. 
Two possible schematic views of this situation are shown in 
Fig.~1. However, one has to be careful with such schematic pictures, 
since these considered convex sets are high-dimensional
and a two-dimensional plot can never characterize all features. 
But, as we will see in this paper, one can derive some 
general statements on the geometry of these sets.

\begin{figure}[t]
\includegraphics[width=36pc]{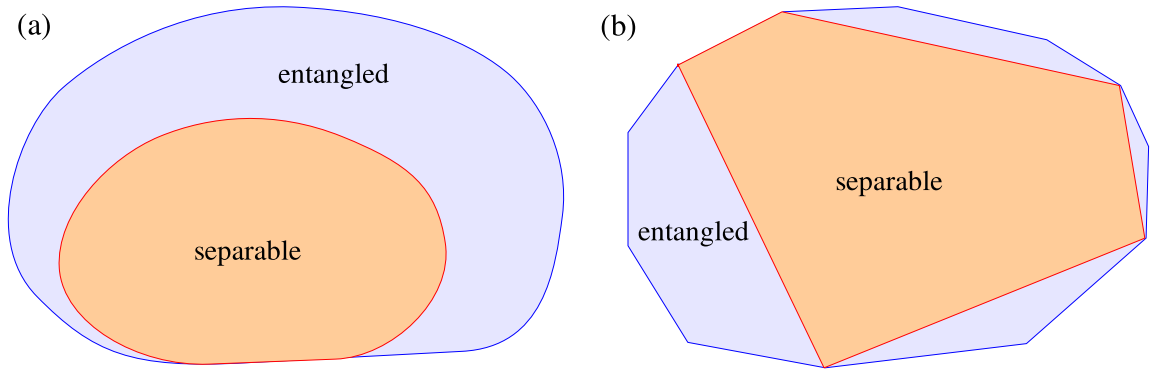}
\caption{
\label{fig1}
(a) Schematic view of the set of separable states as a 
convex subset of the convex set of all states. (b) A 
different view on the same fact. The set of states as 
a polytope with the pure states as corners. Some of 
these corners correspond to product states and they 
span the set of separable states. Note, however, that 
in reality there are infinitely many pure states.}
\end{figure}

The question, whether a given state is separable or entangled 
is the so called separability problem. In order to answer this
question, many separability criteria have been proposed, but 
none of them delivers a complete solution of the problem. 
A famous criterion is the criterion of the positivity of the
partial transposition (PPT criterion, \cite{ppt1}). It is defined 
as follows: given a quantum state $\vr$ in a product basis
\be
\vr=
\sum_{ij,kl} \vr_{ij,kl}
\ket{i}\bra{j}\otimes\ket{k}\bra{l},
\ee
its partial transpose with respect to Bob's system is defined
as
\be
\vr^{T_B}=
\sum_{ij,kl} \vr_{ij,lk} \ket{i}\bra{j}\otimes\ket{k}\bra{l}.
\ee
If $\vr$ is a separable state and has a representation as in 
Eq.~(\ref{sepdef}) it can be easily seen that the partial 
transpose is a valid state and hence a matrix with only positive 
eigenvalues, i.e., $\vr^{T_B}\geq 0.$ Thus, if for a state the partial transpose 
is not positive ($\vr$ is NPT), the state must be entangled. 
Indeed, it has been shown \cite{ppt2} that for $2 \times 2$
and $2 \times 3$ systems a state is PPT if and only if it is 
separable, while for other dimensions there are also PPT 
entangled states. Besides the PPT criterion there are, however,
many other criteria, which may detect a state if the PPT 
criterion fails \cite{ccn1, ccn2, cmc}.

A different approach to the separability problem uses 
entanglement witnesses \cite{ppt2, terhal, optimization, altpra}. 
As already mentioned, 
these are  observables 
with a positive mean value on all separable states, so a 
negative expectation value guarantees that the state is entangled.
Geometrically, witnesses can be seen as hyperplanes which 
separate some entangled states from the set of separable 
states (see Fig.~2 (a)). Here, the hyperplane, indicated as a line 
corresponds to the states with $\mean{\WW}= Tr(\WW\vr)=0.$

\begin{figure}[t]
\includegraphics[width=36pc]{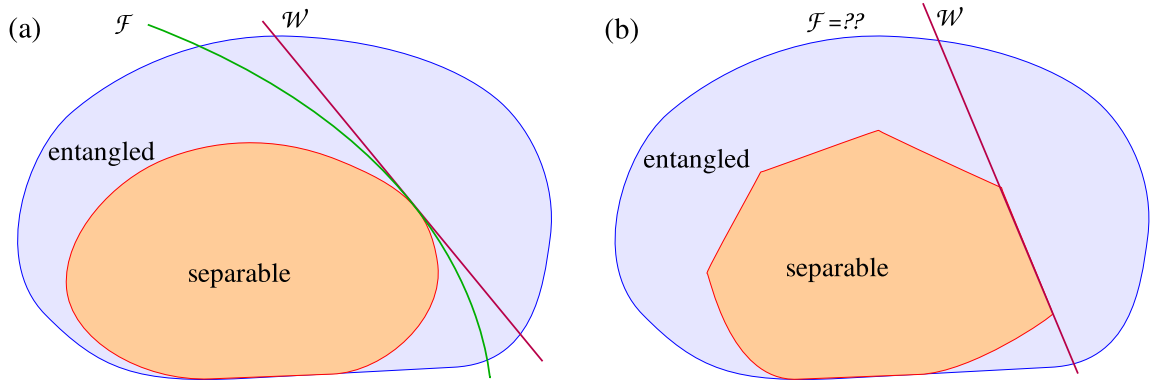}
\caption{
\label{fig2}
(a) Schematic view of an entanglement witness. The line 
$\WW$ denotes the states where $Tr(\vr \WW) =0.$ The curve
$\FF$ sketches a possible nonlinear witness.
(b) If the set of separable states has some facets, it is not 
clear how to find a nonlinear witness. See discussion in the text.}
\end{figure}

{From} the fact that the set of separable states is convex, it 
can easily be deduced that for any entangled state there exist 
a witness detecting it. Finding such a witness, however, is not 
easy, since, as already mentioned, the separability problem is 
not solved yet. But if a state violates a certain separability 
criterion witnesses can often be directly constructed.

To give an example, for any NPT state $\vr_0$ we 
find that $\vr_0^{T_B}$ has a negative eigenvalue 
$\lambda_-$  and a corresponding eigenvector $\ket{\phi}.$ 
Now, an entanglement witness for this state
is given by
\be
\WW = \ketbra{\phi}^{T_B}.
\label{wit1}
\ee
Indeed, due to the identity $Tr(XY^{T_B})=Tr(X^{T_B}Y)$ which 
holds for arbitrary matrices $X,Y$ we have 
$Tr(\vr_0\WW)= Tr(\vr_0^{T_B} \ketbra{\phi}) = \lambda_- < 0$
while for separable (and hence PPT) states we have
$Tr(\vr\WW)=\bra{\phi}\vr^{T_B}\ket{\phi}\geq 0$. 
For the following discussion it is important to note that 
the witness in Eq.~(\ref{wit1}) is not specific:
since the PPT criterion is necessary and sufficient for low
dimensions, witnesses of this type suffice to detect all states
in these systems.  Furthermore, such witnesses can be shown to
be optimal \cite{optimization}, i.e. there is no linear 
witnesses which detects the same states as $\WW$ and some 
states in addition.

In view of Fig.~2(a) it is now a natural question to ask
whether one can improve a linear witness by some nonlinear 
functional. Generically, a witness gives rise to a linear
functional $ \FF_l(\vr)=Tr(\WW \vr) $ and a state $\vr$
is detected whenever $\FF_l(\vr) < 0.$ The aim would be  
to find a nonlinear functional $\FF_{nl} $ of the type
\begin{eqnarray}
\FF_{nl}(\vr) = Tr(\WW \vr) - \NL(\vr),
\end{eqnarray}
which still should be positive on all separable states.
Since we are looking for experimentally implementable
entanglement conditions,  the nonlinearity $\NL(\vr)$
should be a function of some expectation values of
observables. As we will see, one can take a quadratic
polynomial of certain expectation values. It is reasonable 
to consider only $\FF_{nl}(\vr)$ which are stronger
than the linear $\FF_l(\vr)$. So we require that
$\FF_{nl}(\vr)$ detects all the states that are detected
by $\FF_l(\vr),$ and some states in addition.

However, in view of Fig.~2(b) is is not so clear, that 
any witness can be improved by some quadratic terms. 
Indeed, it might be that certain surfaces of the set of 
separable states are not curved, and hence some witnesses
cannot be improved in the way described above. To investigate 
this phenomenon, we need some more terminology \cite{ionou}. 

For a given observable $A$ we call the set 
$H_{A,a}= \{\vr: Tr(A\vr) \leq a\}$ a half-space. For instance, 
the states detected by a witness $\WW$ form just the half-space 
$H_{\WW, 0}.$ A boundary $\pi_{A,a}=\{\vr: Tr(A\vr) = a\}$
is called a hyperplane. 
Further, let $F$ be subset of  a compact convex set $D$
in a $d$-dimensional space. We call $F$ a face, if there 
exists a half-space $H_{A,a}$ with $D\subseteq H_{A,a}$
and $F = D \cap \pi_{A,a}$ for some $A$ and $a.$ 
Note that in this definition $H_{A,a}$ is usually not 
unique. Geometrically, a face is just a set of points at 
the border of $D,$ where all points lie in some hyperplane. 
If a face is of the maximal dimension $d-1,$ we call the face 
a {\it facet.} Then the  half-space $H_{A,a}$ is unique.

Concerning separability, it has been shown in Ref.~\cite{ionou}
that the set of separable states is not a polytope. That is, 
for a description of its borders it is not sufficient to consider 
a {\it finite} number of hyperplanes. One can also say the the border 
does not consist of facets only, hence it must be curved at some points.
The connection between facets and nonlinear 
witnesses becomes also clear: if the set of separable states 
has a facet $F$ as in Fig.~2(b), then this facet corresponds to
unique $H_{A,a}.$ In other words, it defines a witness $\WW.$ 
This witness now can not be improved by a correction like
\be
\FF(\vr)= \mean{\WW} - \mean{X}^2
\ee
for some hermitian $X:$ Since $F$ has dimension $d-1$ and 
$\FF(\vr) = 0$ for all $\vr \in F$ this implies that 
$X =\alpha \WW$ for some real $\alpha.$ Hence, $\FF(\vr)$
does not detect any state in addition to $\WW.$ So the 
existence of facets on the set of separable states is closely 
related  to the existence of nonlinear witnesses.

\section{Basic results on nonlinear witnesses and their 
geometrical interpretation}

Let us now explain the method of Ref.~\cite{wir} for the 
construction of nonlinear witnesses. We consider first
the witness from Eq.~(\ref{wit1}).

As a starting point note that a functional like 
$F(\vr)=\mean{X}\mean{X^\dagger}$ for any $X$ is 
convex in the state. Convexity means that if 
$\vr=\sum_k p_k \vr_k$ is a convex combination 
of some states, then $F(\vr) \leq \sum_k p_k F(\vr_k).$
This fact can be directly calculated, 
see. e.g. \cite{lurs}. Consequently, it implies
that a functional like  
$\GG = \mean{A} - \sum_i \alpha_i \mean{X_i}\mean{X_i^\dagger}$
with $\alpha_i \geq 0 $ is concave in the state. So for
convex combinations we have
$\GG(\vr) \geq \sum_k p_k \GG(\vr_k).$

Let us assume that we have taken 
$ A= \ketbra{\phi}, X_i = \ket{\phi}\bra{\psi_i}$ for 
an arbitrary $\ket{\psi_i}$ and take a separable state 
$\vr.$ Then the partial transpose of the state $\vr^{T_B}$
is again separable and can be written as a convex combination 
of product states, $\vr^{T_B}=\sum_k p_k \ketbra{a_k b_k}.$
For a product vector $\ket{\chi}=\ket{a_k b_k}$ 
we have
\bea
\GG(\ketbra{\chi}) 
&=&  
\braket{\chi}{\phi}\braket{\phi}{\chi}
\cdot
\big[
1 - \sum_i \alpha_i \braket{\chi}{\psi_i}\braket{\psi_i}{\chi}
\big]
\nonumber
\\
&=:&\braket{\chi}{\phi}\braket{\phi}{\chi}
\cdot P(\chi)
\label{nl1}
\eea
Thus, if the function $P(\chi)$ is positive on all
product states, $\GG$ is positive on all product states.
Then, by concavity, it is also positive on convex combinations 
thereof, hence $\GG$ is positive on all separable $\vr^{T_B}$. 
Consequently, with the chosen $X_i,$
\be
\FF(\vr) = \GG(\vr^{T_B})= \mean{\ketbra{\phi}^{T_B}}
- \sum_i \alpha_i \mean{X_i^{T_B}}\mean{(X_i^{T_B})^\dagger}
\label{nl2}
\ee
is positive on all separable states and is hence
a nonlinear improvement of the witness 
$\WW=\ketbra{\phi}^{T_B}.$ From this we have:

{\bf Theorem 1.} 
{\it 
(a) Let $\WW = \ketbra{\phi}^{T_B}$ be an entanglement witness. 
We define $X_i=\ket{\phi}\bra{\psi}$ for an arbitrary $\ket{\psi}$ 
and choose $s(\psi)$ as the square of the largest Schmidt coefficient 
of $\ket{\psi}.$
Then
\be
\FF^{(1)}(\vr)=
\mean{\ketbra{\phi}^{T_B}} 
-
\frac{1}{s(\psi)}
\mean{X^{T_B}} \mean{(X^{T_B})^\dagger}
\label{nl3}
\ee
is a nonlinear improvement of $\WW.$
\\
(b) If we take the same $\WW$ and define 
$X_i=\ket{\phi}\bra{\psi_i}, \;\; i=1,...,K$
with some orthonormal basis $\ket{\psi_i},$ then
\be
\FF^{(2)}(\vr) = \mean{\ketbra{\phi}^{T_B}}
- \sum_{i=1}^K \mean{X_i^{T_B}} \mean{(X_i^{T_B})^\dagger}
\label{nl4}
\ee
is also a nonlinear witness which improves $\WW.$}

{\it Proof.}
(a) In order to show this, note that the squared 
overlap between a state $\ket{\psi}$ and a product 
state is bounded by the maximal squared Schmidt
coefficient \cite{mohamed}, that is, 
$|\braket{\psi}{\chi}|^2 \leq s(\psi).$
From this it directly follows that
$P(\chi)$ in Eq.~(\ref{nl1}) is positive.
(b) For the  $\ket{\psi_i}$  we have in Eq.~(\ref{nl1})
$\sum_i \braket{\chi}{\psi_i}\braket{\psi_i}{\chi}
= Tr(\ketbra{\chi}) = 1$  which  implies that $P(\chi)=0$ and 
proves the claim.
$\qed$

To give an example how this looks like, let us consider the 
two-qubit case. We take $\WW = \ketbra{\phi}^{T_B}$ with 
$\ket{\phi}=(\ket{00}+\ket{11})/\sqrt{2}.$ This witness
can be written as 
\be
\WW = \frac{1}{4}
\big( 
\eins \otimes \eins  
+ \sigma_x \otimes \sigma_x
+ \sigma_y \otimes \sigma_y
+ \sigma_z \otimes \sigma_z
\big).
\ee
This representation shows that $\WW$ can be evaluated by measuring
$\sigma_x \otimes \sigma_x, \sigma_y \otimes \sigma_y$ and
$\sigma_z \otimes \sigma_z,$ and it can be shown that these 
three measurements are the optimal ones \cite{altpra}.
To improve the witness, we take 
$\ket{\psi}=(\ket{01}+\ket{10})/\sqrt{2},$ then a direct calculation
using Theorem 1(a) leads to the nonlinear witness
\be
\FF^{(1)}(\vr) = \mean{\WW} - 
\frac{1}{8} \mean{\sigma_x \otimes \eins + \eins \otimes \sigma_x}^2
- \frac{1}{8} \mean{\sigma_y \otimes \sigma_z - \sigma_z \otimes \sigma_y}^2.
\label{example1}
\ee
If we consider Theorem 1(b) and take the $\ket{\psi_i}$ as the four 
Bell states, we arrive at
\bea
&& \FF^{(2)}(\vr)= \mean{\WW} - 
\frac{1}{16} \Big(\mean{\sigma_x \otimes \eins + \eins \otimes \sigma_x}^2
+ \mean{\sigma_y \otimes \sigma_z - \sigma_z \otimes \sigma_y}^2
+ \mean{\sigma_y \otimes \eins + \eins \otimes \sigma_y}^2
\nonumber
\\
&& \;\;\;\;\;\;\;\; + \mean{\sigma_x \otimes \sigma_z - \sigma_z \otimes \sigma_x}^2
+ \mean{\sigma_z \otimes \eins + \eins \otimes \sigma_z}^2
+ \mean{\sigma_x \otimes \sigma_y - \sigma_y \otimes \sigma_x}^2
+ {\mean{\WW}}^2
\Big).
\label{example2}
\eea
Interestingly, the values of some of the quadratic terms can be determined 
already from the measurements like $\sigma_x \otimes \sigma_x$ etc.~which 
were already needed to evaluate $\WW.$ Hence, the nonlinear witness can be 
used to improve the entanglement detection from the same data given. One 
should also mention that the structure of the nonlinear improvements as a sum 
of squares is generic: we can write the term $X^{T_B}=H+i \cdot A$ as a 
sum of its hermitian and anti-hermitian part, where $H$ and $A$ are 
hermitian. Then we have $\mean{X^{T_B}} \mean{(X^{T_B})^\dagger}
= \mean{H}^2+ \mean{A}^2,$  which leads to this structure.

Note that Theorem 1 provides a whole class of nonlinear 
improvements, since one can pick an arbitrary $\ket{\psi}$ and 
compute the corresponding nonlinearity. This freedom may be used 
to design nonlinear witnesses for special experimental purposes.
Concerning the strength of the nonlinear improvements for the case 
of two qubits it has been shown in  Ref.~\cite{wir} that nonlinear 
witnesses like Eq.~(\ref{example1}, \ref{example2}) improve the 
witness $\WW$ quite significantly. For the general case, we state
a result from Ref.~\cite{wir} without the proof:

{\bf Theorem 2.}
{\it 
(a) Let $\WW = \ketbra{\phi}^{T_B}$ be a witness. A state $\vr$ 
can be detected by a witness of the type $\FF^{(1)}$ from 
Eq.~(\ref{nl3}) if and only if
\be
\bra{\phi}\vr^{T_B}\ket{\phi} <
\Big[
Tr_B
\big(
\sqrt{Tr_A(\vr^{T_B}\ketbra{\phi}\vr^{T_B})}
\big)
\Big]^2
.
\label{nl5}
\ee
\\
(b) In the same situation, a state $\vr$ can be detected by 
a witness of the type $\FF^{(2)}$  from Eq.~(\ref{nl4}) 
if and only if
\be
\bra{\phi}\vr^{T_B}\ket{\phi} < \bra{\phi}(\vr^{T_B})^2 \ket{\phi}
\label{nl6}
\ee
holds. In this case, the state is detected by all nonlinear witnesses
of the type $\FF^{(2)}.$ 
\\
(c) Finally, if Eq.~(\ref{nl6}) is fulfilled, then also Eq.~(\ref{nl5}) 
holds, thus the witnesses of the type $\FF^{(1)}$ are stronger.
Furthermore, Eqs.~(\ref{nl5}, \ref{nl6}) are never fulfilled for
PPT states.
}

One interesting point in this Theorem is the fact that the 
nonlinear improvements do not detect PPT states. The witness
$\WW$ is derived  from the separability criterion of the positivity
of the partial transpose and the nonlinear improvements of $\WW$ 
are not more powerful than the original PPT criterion. This may sound 
disappointing at first sight. One should note, however, that the proof 
relies on special results for the PPT criterion and it is unlikely that
the same fact holds also for witnesses derived from other entanglement 
criteria.

Let us now discuss nonlinear improvements for other
entanglement witnesses, which are not related to the PPT
criterion. This can be done via the theory of {positive maps.}
Let us shortly explain this subject. Let $\HH_B$ and $\HH_C$ 
be Hilbert spaces and let $\BB(\HH_i)$ denote the linear operators 
on it. A linear map $\Lambda: \BB(\HH_B) \rightarrow \BB(\HH_C)$
is called {positive} if (a) it maps hermitian operators onto 
hermitian operators, fulfilling $\Lambda(X^\dagger)=\Lambda(X)^\dagger$ 
and (b) it preserves the positivity, i.e. if $X \geq 0$ then 
$\Lambda(X) \geq 0.$ Note that the second condition implies 
that it maps valid density matrices onto density matrices (up 
to a normalization). A positive map $\Lambda$ is called {\it completely} 
positive 
when for an arbitrary $\HH_A$ the map $\II_A \otimes \Lambda$ 
is positive, otherwise, it $\Lambda$ is positive, but not 
completely positive. Here, $\II_A$ denotes the identity on $\BB(\HH_A).$
For example, the transposition is positive, but not completely positive: 
while $X \geq 0$ implies $X^T \geq 0,$ the {\it partial} transposition 
does not preserve the positivity of a state.

Thus, similarly as the PPT criterion, other entanglement criteria 
can be formulated from other positive, but not completely positive
 maps. Indeed, it has been shown 
\cite{ppt2, hororeview} that a state $\vr \in \BB(\HH_A) \otimes \BB(\HH_B)$ 
is separable if and only if for all positive maps $\Lambda$ the relation
\be 
\II_A \otimes \Lambda (\vr) \geq 0
\ee
holds. Consequently, if $\vr$ is entangled there must
be a positive, but not completely positive map $\Lambda$ 
where $\II_A \otimes \Lambda (\vr)$ has a negative 
eigenvalue $\lambda_-$ and a corresponding eigenvector 
$\ket{\phi}.$ Taking $(\II_A \otimes \Lambda)^+$  as the 
adjoint of the map $(\II_A \otimes \Lambda)$ with respect
to the scalar product $\braket{X}{Y}=Tr(X^\dagger Y)$ a witness
detecting $\vr$ is given by
\be
\WW= (\II_A \otimes \Lambda)^+ (\ketbra{\phi}),
\label{nl14}
\ee
since we have 
$Tr[\rho \WW] = Tr[\rho (\II_A \otimes \Lambda)^+(\ketbra{\phi})]
= Tr[\II_A \otimes \Lambda (\vr) \ketbra{\phi}] = \lambda_-.$ 
By some rescaling we can always achieve that $(\II_A \otimes \Lambda)^+$ 
is not trace increasing. Then this witness can be improved as shown 
in Theorem 1.

Starting with an arbitrary witness, we make use of the Jamio{\l}kowski 
isomorphism \cite{hororeview, jamiol} between operators
and maps. According to this, an operator $E$ on
$\BB(\HH_B) \otimes \BB(\HH_C)$ corresponds to a map
$ \varepsilon : \BB(\HH_B) \rightarrow \BB(\HH_C)$
acting as  
\be
\varepsilon (\vr) = Tr_B (E \vr^T \otimes \eins_C).
\label{nl15}
\ee
Conversely, we have
\be
E = (\II_{B'} \otimes \varepsilon)(\ketbra{\phi^+}),
\ee
where $\HH_{B'}\cong \HH_{B}$ and $\ket{\phi^+}= \sum \ket{ii}$
is a maximally entangled state on  $\HH_{B'} \otimes \HH_B.$
The important fact is that if $E$ is an entanglement witness, then
$\varepsilon$ is a positive, but not completely positive map 
\cite{hororeview}. Again, by rescaling the witness
we can achieve that the positive map is not trace 
increasing. Hence, any witness can be written as in Eq.~(\ref{nl14}) 
for a suitable positive map and we arrive at:

{\bf Theorem 3.}
{\it 
Any bipartite entanglement witness can be improved
by nonlinear corrections. This can be done by 
calculating the corresponding 
positive map from the Jamio{\l}kowski 
isomorphism and then applying the methods of Theorem 1.
}

According to our discussion in the previous Section, we 
can directly conclude:
 
{\bf Theorem 4.} 
{\it The set of separable states has no facets.}

The generic construction of nonlinear witnesses allows to 
conclude that at the border between  separable and  
entangled states there is no facet. But also at the border 
between the separable states and the non-positive matrices
there are no facets: any facet of this kind would correspond 
to a $\WW$ which is positive. Then, in Eq.~(\ref{nl15}) the 
map $\varepsilon$ is completely positive, but nevertheless 
one can derive from it a nonlinear functional which is 
positive on all separable states as in Theorem 1.

Note that the Theorem 4 does not imply that the surface of the 
set of separable states is a manifold which is differentiable
in every 
point. It still may have some edges or faces, however, these 
edges do not have the maximal possible dimension.

\section{An alternative derivation}

In this Section, we give an alternative way of deriving 
nonlinear improvements for a given linear witness. This 
proof  uses covariance matrices for the construction. 
Although the derivation is completely different,  
the resulting nonlinear witnesses are similar to the 
constructions of the previous section, they are, however, 
slightly weaker. The method presented here  
is closely related to the results of Ref.~\cite{vogel, rigas, piani}, 
and may be used to improve some separability criteria given in 
these references.

To start, let $A_k$ be a basis of the operator space for 
Alice, and let $B_k$ be a basis of the operator space for 
Bob. That is, for two qubits the $A_k, B_k$ may be the 
Pauli matrices including the identity. For a given state $\vr$ 
we define a hermitian matrix
$\eta$  with the entries $\eta_{i;j}=\eta_{i_1,i_2;j_1,j_2}$ 
via
\be
\eta(\vr, A_k, B_k) \equiv \eta_{i_1,i_2;j_1,j_2} 
:= \mean{A_{i_1}A_{j_1}\otimes B_{i_2}B_{j_2}}.
\ee
The notion $\eta(\vr, A_k, B_k)$ should emphasize the 
dependence of $\eta$ on the state and the observables, 
and $\eta_{i_1,i_2;j_1,j_2}$ emphasizes the entries of 
$\eta.$ We mix both notations when there is no risk of 
confusion.
We then define the partial transposition of $\eta$ as
\be
\eta^{T_B} = \eta_{i_1,j_2;j_1,i_2} := 
\mean{A_{i_1}A_{j_1}\otimes B_{j_2}B_{i_2}}.
\ee 
Then we have the following Lemma.

{\bf Lemma 5.} 
{\it 
(a) We have always $\eta \geq 0,$ i.e., $\eta$ is a positive matrix.
\\
(b) The partial transposition fulfills
\be
\eta^{T_B}(\vr, A_k, B_k) = \eta(\vr^{T_B}, A_k, B^T_k)
\ee
(c) For a state $\vr$ we have $\eta^{T_B}\geq 0$ if and only if 
$\vr$ is PPT.
}

{\it Proof.} (a) It is known that the (asymmetric) covariance matrix
\be
\gamma= \mean{A_{i_1}A_{j_1}\otimes B_{i_2}B_{j_2}} -
\mean{A_{i_1}\otimes B_{i_2}}\mean{A_{j_1}\otimes B_{j_2}}
\ee
is always positive semidefinite \cite{robertson}. The nonlinear
part $\mean{A_{i_1}\otimes B_{i_2}}\mean{A_{j_1}\otimes B_{j_2}}$
of it is also positive (and subtracted), thus the linear part, which 
corresponds to $\eta,$ must be positive. (b) This can be simply 
calculated, using the general fact that $Tr(X^{T_B} Y)= Tr(X Y^{T_B}).$
(c) The direction ``$\Leftarrow$'' follows already from (a) and (b).
To see the other direction, assume that $\vr^{T_B}\not \geq 0.$
Then, $\vr^{T_B}$ must have some negative eigenvalue $\lambda_-$ 
and a corresponding eigenvector $\ket{\phi}.$ Now we can expand the 
operator 
$\ketbra{\phi}= \ketbra{\phi}^2 = \sum_{i_1,j_1} \alpha_{i_1, i_2} A_{i_1} \otimes B^T_{i_2}$
in the operator basis. 
Then we have 
\be
\lambda_- = Tr(\vr^{T_B} \ketbra{\phi}) = 
\sum_{i_1,i_2} \sum_{j_1,j_2}  
\alpha_{i_1, i_2} 
\eta^{T_B}(\vr, A_k, B_k)_{i_1,i_2;j_1,j_2}
\alpha_{j_1, j_2}  = \bra{\alpha} \eta^{T_B} \ket{\alpha}< 0,
\ee
as can be checked by direct calculation. This proves the claim.
$\qed$

Before we can improve witnesses, we need one more definition. 
We define:
\bea
\chi &:=& \mean{A_{i_1}\otimes B_{i_2}}\mean{A_{j_1}\otimes B_{j_2}},
\\
\Gamma&:=& \eta^{T_B}-\chi.
\eea
$\chi$ is just the nonlinear part of the covariance matrix. $\Gamma$
is similar to the covariance matrix, but in general it is not a 
covariance matrix. However, if $\vr$ is PPT, then $\Gamma$ is the 
covariance matrix for the observables ${A_{i_1}\otimes B^T_{i_2}}$
in the state $\vr^{T_B}.$ Thus, in this case it is also positive.

Let us assume that we have an NPT state $\vr$  and consider a 
witness $ \WW = \ketbra{\phi}^{T_B}$ as in Eq.~(\ref{wit1}).
More generally, we could also consider a witness of the type 
$P^{T_B}$ where $P$ is a positive operator.
If we find a positive operator $Q = Q_{i_1,i_2;j_1,j_2} \geq 0$ 
such that
\be
\ketbra{\phi} = \sum_{i_1,i_2} \sum_{j_1,j_2}  
Q_{i_1,i_2;j_1,j_2} A_{i_1}A_{j_1}\otimes B^T_{i_2}B^T_{j_2},
\label{qdef}
\ee
we have 
\be
Tr(\Gamma Q) \geq 0
\ee
for all PPT states. However, we have also
\be
Tr(\Gamma Q) = \ketbra{\phi}^{T_B} - Tr(\chi Q). 
\ee
This implies that $Tr(\Gamma Q)$ is the desired nonlinear 
functional which improves the witness $\WW.$

The question remains, whether such a $Q$ can always be found. 
Indeed this is the case. We can always construct it as above in
the proof of the Lemma 5 (c). This construction, however, is not 
very useful, since it leads to nonlinear functionals of the type
$\FF=\mean{\WW} - \mean{\WW}^2$ which are not better than the 
witness. But we can choose other $Q,$ since the observables
$A_{i_1}A_{j_1}\otimes B^T_{i_2}B^T_{j_2}$ are tomographically 
overcomplete and thus $Q$ is  by no means unique.
The characterization of the possible $Q$ can be summarized
as follows:

{\bf Theorem 6.}
{\it 
(a) The set of possible $Q$ is closed and convex. 
\\
(b) For entanglement detection it suffices to consider
the extremal points of this set. If $Q$ is of rank one, 
it is extremal.
\\
(c) {All} pure  extremal points can be found as follows: 
For the given $P\geq 0$  (e.g. $P=\ketbra{\phi}$)
one considers the spectral decomposition of $\sqrt{P}$, that 
is
$
\sqrt{P}= U D U^\dagger
$
with $U$ unitary and $D$ diagonal. Then one considers
\be
X =  U D V = \sum_{i_1,i_2}  
\alpha_{i_1 i_2} A_{i_1} \otimes B^T_{i_2},
\label{xdef}
\ee
with arbitrary unitary $V$ and thus complex $\alpha_{i_1 i_2}.$
One extremal $Q$ is then given by 
\be
Q_{i_1,i_2;j_1,j_2} = \alpha_{i_1 i_2} \alpha^*_{j_1 j_2}.
\label{qextrem}
\ee
More generally, the search for an appropriate $Q$ can be solved via
the semidefinite program \cite{sdp}
\bea
\mbox{\rm minimize} && Tr(\Gamma Q), 
\\
\mbox{\rm subject to}&& Q \geq 0,
\\ &&  P = \sum_{i_1,i_2} \sum_{j_1,j_2}  
Q_{i_1,i_2,j_1,j_2} A_{i_1}A_{j_1}\otimes B^T_{i_2}B^T_{j_2}.
\eea  
(d) The detection power of the resulting nonlinear separability 
criteria does not depend on the initial choice of the $A_k$ and 
$B_k.$
}

{\it Proof.} (a) This is obvious. (b) It is clear, that the 
extremal points suffice, since the resulting entanglement 
conditions are linear in $Q.$ (c) It can be straightforwardly 
seen that the constructed $Q$ are valid: We have 
$P=\sqrt{P}\sqrt{P} = X X^\dagger$ from this it follows that 
Eq.~(\ref{qdef}) holds. On the other hand, any pure extremal 
$Q$ must be of the type Eq.~(\ref{qextrem}). This implies 
that we can find a corresponding $X$ with $X X^\dagger =P$ 
of the desired type. 
Here, we use that the singular value decomposition is 
unique. (d) Assume that we take other observables like 
$
\tilde A_k = \sum_l C_{kl}  A_l $ and $\tilde B_k = \sum_l D_{kl}  B_l.$
Then the matrices $C,D$ must be invertible and  we have
\bea
\eta(\vr, \tilde A_k,\tilde B_k) &=& C \otimes D \; \eta(\vr, A_l, B_l) C^T \otimes D^T,
\\
\Gamma(\vr, \tilde A_k,\tilde B_k)&=& C \otimes D \; \Gamma(\vr, A_l, B_l) C^T \otimes D^T.
\eea
From Eq.~(\ref{xdef}) one can read off that $\alpha$ transforms like
$C^T \otimes D^T \alpha (\tilde A_k,\tilde B_k) =  \alpha (A_l, B_l),$ 
this implies that
$Q(\tilde A_k,\tilde B_k) = (C^T \otimes D^T)^{-1}Q(A_l,B_l) (C \otimes D)^{-1}.$
This proves the claim. 
$\qed$

{\bf Corollary 7.} 
{\it Any entanglement witness of the form $\WW=\ketbra{\phi}^{T_B}$ 
can be improved by some quadratic corrections.}

{\it Proof.} We only have to show that the above given improvements 
are not {all} trivial. Let us assume the contrary. This would imply 
that for all states $\vr$ with $Tr(\WW \vr) = 0$ and for all possible 
$Q$ we would have
\be
\sum_{i_1,i_2} 
\sum_{j_1,j_2}  
Q_{i_1,i_2;j_1,j_2} 
\mean{A_{i_1}\otimes B_{i_2}} \mean{A_{j_1}\otimes B_{j_2}} = 0.
\label{contra}
\ee
Defining $\beta_{i_1,i_2}=\mean{A_{i_1}\otimes B_{i_2}}$ this may be 
written as $\bra{\beta}Q\ket{\beta}=0.$ For a $d\times d$ system the 
set of all density matrices is a $d^2 \times d^2 - 1$ (real) dimensional 
manifold. The set of  states $\vr$ with $Tr(\WW \vr) = 0$ forms a 
$d^2 \times d^2 - 2$ dimensional affine space $F.$ Consequently, the 
possible $\ket{\beta}$ arising from the $\vr \in F$ span a 
$d^2 \times d^2 - 1$ dimensional subspace.
Since $Q$ is a $d^4 \times d^4$ matrix it has $d^4$ eigenvectors and 
nonnegative eigenvalues. So if  Eq.~(\ref{contra}) were valid, then 
$d^4-1$ of the eigenvalues would equal zero, hence all $Q$ would be of 
rank one, and all valid $Q$ would be a multiple of a fixed projector. 
But obviously there are more than one valid 
$\alpha_{i_1,i_2}$ in Eqs.~(\ref{xdef}, \ref{qextrem}).
$\qed$

This approach gives a different view at nonlinear entanglement witnesses. 
As in the previous section, it can be directly extended to other positive maps
besides the partial transposition. For a witness of the type $\WW=\ketbra{\phi}^{T_B}$
the construction in Theorem 6 would first start with a choice of a $\ket{\psi}$ to 
build up  $X=\ket{\psi}\bra{\phi}$ and finally one arrives at the nonlinear witness
as in Theorem 1(a), but without the factor $1/s(\psi).$ Since this prefactor is 
always larger than one, the witness from Theorem 1 is stronger. 

\section{The multipartite setting}

Finally, let us discuss shortly the extension to the multipartite 
case. We concentrate on the three-qubit case, since this already 
suffices to state the main results. 

First, it is important to note that for three parties several forms of entanglement 
exist \cite{acin}. 
We call a tripartite  pure state fully separable if it is of the form
$\ket{\psi}=\ket{a}\otimes \ket{b}\otimes\ket{c},$ and a mixed state 
is fully separable, if it is of the form 
\be
\vr = \sum_i p_i \ketbra{a_i}\otimes \ketbra{b_i}\otimes\ketbra{c_i}.
\ee
Furthermore, a pure state is biseparable, if it is separable with respect 
to one of the three possible bipartitions $A|BC,$ $AB|C$ or $AC|B,$ 
e.g.~$\ket{\psi}=\ket{\phi}_{AB} \otimes \ket{\chi}_C.$ Again, a mixed state is 
called biseparable, if it can be written as a convex sum of biseparable states, 
\be
\vr = \sum_i p_i \ketbra{\psi^{\rm (bs)}_i},
\label{bisep}
\ee
where the biseparable states $\ket{\psi^{\rm (bs)}_i}$ might be biseparable 
with respect to different partitions. Otherwise, the state $\vr$ is called 
genuine multipartite entangled. 

For these types of entanglement one can define entanglement witnesses as for the 
bipartite case. However, it is important to note  that in 
experiments mainly genuine multipartite entanglement is of interest: to 
confirm the success of an experiment with three qubits one has to show that 
all three qubits were entangled, and not only two of them. Hence it is 
not sufficient to exclude full separability.

The question arises, whether we can construct nonlinear entanglement witnesses 
also for the multipartite case. In one case this can be done. 
Let $\WW$ by a witness ruling full separability, that is, $Tr(\vr \WW) \geq 0$
for all fully separable states. Then we can pick the bipartition $A|BC$ and 
find a nonlinear improvement $\FF_{A|BC}(\vr)$ for this bipartition. Then,  
the minimum over all bipartitions,
\be
\FF_{\rm tot}(\vr) = \min \{ \FF_{A|BC}(\vr),\FF_{AB|C}(\vr),\FF_{AC|B}(\vr) \},
\label{fullsepwit}
\ee
is clearly positive on all fully separable states. Hence, it defines a nonlinear 
improvement of $\WW.$

For the more interesting case of witnesses for genuine multipartite entanglement
this recipe does not work anymore. Here, the first problem comes from the fact 
that in the definition of biseparability in Eq.~(\ref{bisep}) also convex 
combinations of biseparable states with respect to different bipartitions 
are allowed. One might be tempted to consider in analogy to Eq.~(\ref{fullsepwit})
a function like 
$\FF_{\rm tot}(\vr) = \max \{ \FF_{A|BC}(\vr),\FF_{AB|C}(\vr),\FF_{AC|B}(\vr)\}$
to improve a  witness for genuine multipartite entanglement. While this functional
is positive on all pure biseparable states, it is however, not necessarily
positive on mixed biseparable states, since the maximum of concave functions is 
not concave.

Another attempt for the improvement of witnesses for genuine multipartite 
entanglement is to calculate the $\FF_{A|BC}(\vr)$ etc.~as above, and then 
one can investigate the quadratic terms for the $\FF$ as in 
Eqs.~(\ref{example1}, \ref{example2}). If one finds a quadratic term which occurs
in all $\FF,$ then this term may be subtracted from $\WW,$ arriving at a valid 
witness. However, it seems quite difficult to find a single example where this 
recipe works. So the search for nonlinear entanglement witnesses for multipartite
systems remains an interesting and open problem for further study. 

\section{Conclusion}

In conclusion we investigated nonlinear entanglement witnesses from 
different perspectives. We demonstrated that they can be used to show 
that the set of separable states has no facets. We also gave a new 
derivation of nonlinear witnesses based on covariance matrices. This 
highlights the close connection between nonlinear entanglement detection 
and separability criteria in terms of covariance matrices.
Finally, we discussed the problems which occur if one wishes to construct
nonlinear witnesses for the multipartite case.

\ack
We thank Tobias Moroder for discussions and the organizers of DICE2006
for their work. 
This work was supported by the FWF and the EU (OLAQI, QAP, QICS, SCALA).

\end{document}